# Silicon Implantation and Annealing in $\beta$-Ga$_2$O$_3$: Role of Ambient, Temperature, and Time

Katie R. Gann[1, a)], Naomi Pieczulewski[1], Cameron A. Gorsak[1], Karen Heinselman[2], Thaddeus J. Asel[3], Brenton A. Noesges[3,4], Kathleen T. Smith[5], Daniel M. Dryden[6], Huili Grace Xing[1,7,8], Hari P. Nair[1], David A. Muller[5,8], Michael O. Thompson[1]

[1]Department of Materials Science and Engineering, Cornell University, Ithaca, New York, 14853, USA
[2]National Renewable Energy Laboratory, Golden, Colorado, 80401, USA
[3]Air Force Research Laboratory, Materials and Manufacturing Directorate, Wright-Patterson AFB, Ohio 45433, USA
[4]Azimuth Corporation, Beavercreek, Ohio, 45324, USA
[5]School of Applied and Engineering Physics, Cornell University, Ithaca, New York, 14853, USA
[6]Air Force Research Laboratory, Sensors Directorate, Wright-Patterson AFB, Ohio 45433, USA
[7]School of Electrical and Computer Engineering, Cornell University, Ithaca, New York, 14853, USA
[8]Kavli Institute at Cornell for Nanoscale Science, Cornell University, Ithaca, New York, 14853, USA

a) Electronic mail: krg66@cornell.edu

Optimizing thermal anneals of Si-implanted $\beta$-Ga$_2$O$_3$ is critical for low resistance contacts and selective area doping. We report the impact of annealing ambient, temperature, and time on activation of room temperature ion-implanted Si in $\beta$-Ga$_2$O$_3$ at concentrations from $5\times10^{18}$ to $1\times10^{20}$ cm$^{-3}$, demonstrating full activation (>80% activation, mobilities >70 cm$^2$/Vs) with contact resistances below 0.29 $\Omega$-mm. Homoepitaxial $\beta$-Ga$_2$O$_3$ films, grown by plasma assisted MBE on Fe-doped (010) substrates, were implanted at multiple energies to yield 100 nm box profiles of $5\times10^{18}$, $5\times10^{19}$, and $1\times10^{20}$ cm$^{-3}$. Anneals were performed in a UHV-compatible quartz furnace at 1 bar with well-controlled gas composition. To maintain $\beta$-Ga$_2$O$_3$ stability, $p_{O2}$ must be greater than $10^{-9}$ bar. Anneals up to $p_{O2}$ = 1 bar achieve full activation at $5\times10^{18}$ cm$^{-3}$, while $5\times10^{19}$ cm$^{-3}$ must be annealed with $p_{O2} \leq 10^{-4}$ bar and $1\times10^{20}$ cm$^{-3}$ requires $p_{O2} < 10^{-6}$ bar. Water vapor prevents activation and must be maintained below $10^{-8}$ bar. Activation is achieved for anneal temperatures as low as 850 °C with mobility increasing with anneal temperature up to 1050 °C, though Si diffusion has been reported above 950 °C. At 950 °C, activation is maximized between 5 and 20 minutes with longer times resulting in decreased carrier activation (over-annealing). This over-annealing is significant for concentrations above $5\times10^{19}$ cm$^{-3}$ and occurs rapidly at $1\times10^{20}$ cm$^{-3}$. RBS (channeling) suggests damage recovery is seeded from remnant aligned $\beta$-Ga$_2$O$_3$ that remains after implantation; this conclusion is also supported by STEM showing retention of the $\beta$-phase with inclusions that resemble the $\gamma$-phase.

## I. INTRODUCTION

Beta-phase gallium oxide ($\beta$-Ga$_2$O$_3$) has received attention in recent years due to its ultrawide band gap (~4.8 eV), estimated high breakdown strength (~ 8 MV/cm), and optical transparency.[1,2] While other metastable polymorphs are also of interest, the monoclinic $\beta$-phase has been extensively studied[3,4] and the availability of large area melt-grown substrates is a distinct advantage over other wide and ultra-wide bandgap semiconductors.[1,5,6] $\beta$-Ga$_2$O$_3$ can be readily doped with a variety of n-type donors including Si, Sn, and Ge with Si emerging as the dopant of choice.[7,8] *In-situ* doping during epitaxial growth has been demonstrated during metalorganic chemical vapor deposition (MOCVD), pulsed laser deposition (PLD) and molecular beam epitaxy (MBE) with Si concentrations up to $2\times10^{20}$ cm$^{-3}$.[9–13] Si n-type doping by ion implantation has also been demonstrated, providing a controllable method for selective area doping in lateral devices.[14–19]

Ion implantation requires thermal annealing to remove implantation-induced lattice damage, including point and extended defects as well as radiation induced phase transformations,[20–23] and to activate implanted dopants. Processing parameters for annealing include time, temperature, heating and cooling rates, ambient conditions, and the presence of a protective layer during annealing. Ion implantation of Si, Ge, and Sn [14,18,24,25] have been reported for n-type doping for channel and contact regions in $\beta$-Ga$_2$O$_3$, while implants of Mg and N have been investigated as deep acceptors for blocking layers.[25,26] Sasaki, in 2013, reported activation of Si after annealing between 900 and 1100 °C. While increasing temperature improved activation, significant Si diffusion was observed at 1100 °C. Sasaki also reported a decrease in activation fraction as the implanted concentration increased from $1\times10^{19}$ to $1\times10^{20}$ cm$^{-3}$ Si.[14] Tadjer, in 2019, studied the lattice recovery after Si and Sn implant at doses of $2\times10^{15}$ cm$^{-2}$, corresponding to peak concentrations of $2\times10^{20}$ cm$^{-3}$. They reported that lattice recovery required anneals at 1150 °C for the highest dose Si implants, with higher temperatures required for Sn implants, consistent with the higher atomic mass implant generating more lattice damage.[24] In 2022, Spencer demonstrated activation of Si, Ga, and Sn implants after annealing at 925 °C for 30 minutes by rapid thermal annealing (RTA), achieving up to 65% activation for the implanted Si, corresponding to $1.3\times10^{19}$ cm$^{-3}$ and a mobility of 93 cm$^2$/V-s.[18] Further, by increasing the temperature to 600 °C during implantation, as compared to room temperature, Sardar demonstrated in 2022 an activation fraction of 82% for Si implanted to a peak concentration of $1.2\times10^{20}$ cm$^{-3}$.[16]

Annealing under N$_2$ ambients has been shown to be favorable compared to O$_2$ ambients, with N$_2$ anneals activating carriers while O$_2$ reversibly deactivates carriers and enhances Si diffusion.[27,28] Annealing in argon has been reported as similar to annealing in N$_2$, suggesting that the inert gas does not affect activation.[29,30] Existing literature does not quantify gas purity, especially trace concentrations of oxygen and water in N$_2$ or Ar ambients. Some $p_{O2}$ is critical for annealing as $\beta$-Ga$_2$O$_3$ is unstable at high temperature in the absence of oxygen, decomposing to the volatile Ga$_2$O sub-oxide or to Ga metal; above 1150 °C, $\beta$-Ga$_2$O$_3$ has been shown to decompose under nominally pure N$_2$ and the addition of H$_2$ lowered the decomposition threshold to 350 °C.[31] Lany estimated the equilibrium partial pressure of Ga$_2$O as a function of temperature and p$_{O2}$,[7] indicating that limiting p$_{Ga2O}$ to < 10$^{-5}$ bar requires that the concentration of O$_2$ must be maintained above 10$^{-14}$ bar at 900 °C, 10$^{-10}$ bar at 1000 °C, and 10$^{-7}$ bar at 1100 °C. Their DFT calculations[7] also suggested a strong p$_{O2}$ dependence for Si activation, especially at high carrier concentrations.

Despite the early successes of Si ion implantation in $\beta$-Ga$_2$O$_3$, detailed studies on the effects of annealing ambient, temperature, and time are absent from the literature. In this work, we report on the activation and mobility following furnace annealing of room temperature Si implants (from 5×10$^{18}$ cm$^{-3}$ to 1×10$^{20}$ cm$^{-3}$) as a function of annealing ambient (controlled p$_{O2}$ and p$_{H2O}$), temperature (850 °C to 1050 °C), and time (2.5 to 120 minutes). Under optimized annealing conditions, activation to >80% with mobilities >70 cm$^2$/V-s was observed for concentrations up to 1×10$^{20}$ cm$^{-3}$.

## II. EXPERIMENTAL

$\beta$-Ga$_2$O$_3$ films were grown using a Veeco GEN Xvel plasma-assisted MBE (PAMBE) system equipped with a standard effusion cell for Ga and a UNI-Bulb RF plasma source angled at 45° relative to the substrate for oxygen. Tamura Novel Crystal Technology (NCT) Fe-doped $\beta$-Ga$_2$O$_3$ (010) substrates (23×25 mm) were solvent cleaned prior to loading into the growth chamber, with an additional *in-situ* oxygen plasma clean prior to growth. Unintentionally doped (UID) films were grown at a substrate temperature of 650 °C with a Ga beam equivalent pressure (BEP) of 6.0×10$^{-8}$ Torr (calibrated to be near the stoichiometric conditions for the chamber), an oxygen plasma from 2.0 sccm of $O_2$ flow, and 250 W of RF power. A target film thickness of 400 nm served as a buffer layer to minimize Fe diffusion from the substrate into the surface implanted layer.[32]

To compare with implanted samples, an *in-situ* doped sample was grown in an Agnitron Agilis 100 MOCVD system on equivalent substrates. A ~50 nm UID layer was first grown at a reactor pressure of 15 Torr and a substrate temperature of 600 °C, followed by growth of a ~95 nm film doped with Si at $6.9 \times 10^{19}$ cm$^{-3}$ grown at a pressure of 40 Torr and a substrate temperature of 705 °C. The substrate temperature was measured using a pyrometer aligned to the backside of the SiC-coated graphite susceptor. Triethylgallium (TEGa) and silane (25 ppm $SiH_4$ in argon), were used as precursors for gallium and silicon, respectively, with argon as the carrier gas and molecular oxygen as the oxidant. The TEGa molar flow was 19 and 39 µmol/min for the UID and doped layer, respectively. For the doped layer, the silane flow was 26.81 nmol/min. For the entire growth, the oxygen flow was set at 500 sccm with a total gas flow of 6000 sccm.

Prior to ion implantation, films were capped with approximately 20 nm $SiO_2$ via atomic layer deposition (ALD) in an Oxford FlexAL system at 300 °C using tris(dimethylamino)silane. Ion implantation was performed using three implant energies to form a 100 nm box-shaped concentration profile with straggle to ~200 nm (Fig. 1). Three Si box concentrations of $5 \times 10^{18}$, $5 \times 10^{19}$, and $1 \times 10^{20}$ cm$^{-3}$ were formed, all using similar implant energies (modified to compensate for slight variations in the ALD thickness). The nominal energies and doses for the $5 \times 10^{19}$ cm$^{-3}$ box implant were: 15, 45, and 115 keV at $5.6 \times 10^{13}$, $1.4 \times 10^{14}$, $5.2 \times 10^{14}$ cm$^{-2}$, respectively. Figure 1 shows the simulated profile from stopping range of ions in matter (SRIM)[33] and measured Si concentration by secondary ion mass spectrometry (SIMS) confirming the box profile. Table SI in supplemental information (SI) gives specific implant details for each growth.

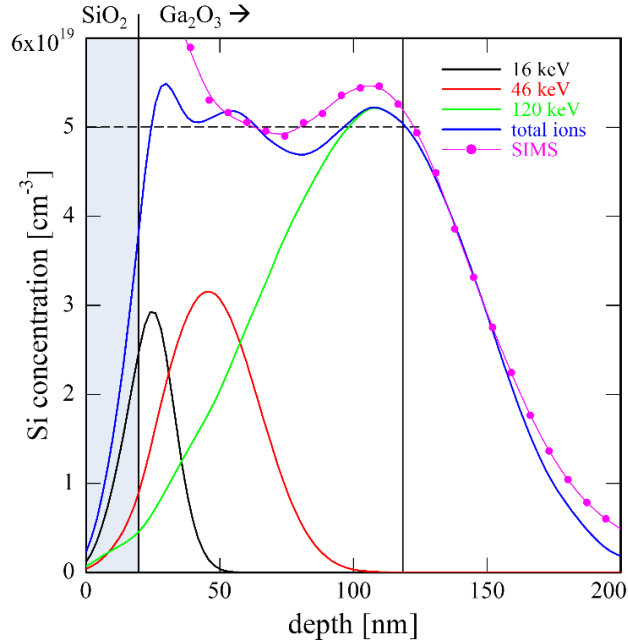

FIG 1. Simulated implant profile for $5\times10^{19}$ cm$^{-3}$ implant with 16, 46, and 120 keV at doses of $6.1\times10^{13}$, $1.4\times10^{14}$, $5.1\times10^{14}$ cm$^{-2}$, respectively, creating a box-shaped implant for the first 100 nm of the film with straggle to 200 nm. SIMS data shown is for the same implant parameters, and data below 30 nm is often unreliable due to potential surface contamination.

After implant, samples were diced into 5×5 mm die for anneals. The SiO$_2$ layer was removed using 6:1 buffered oxide etchant for one minute. To account for small differences in UID growth and substrate quality, all trends reported for different activation conditions only include samples from one growth. While rapid thermal annealing (RTA) is often used to activate implants, furnace annealing was chosen to permit careful control of gas purity, furnace cleanliness, purging times, and temperature accuracy.

Annealing was performed in an ultra-high vacuum (UHV) compatible quartz tube furnace, shown schematically in Figure S1. Gas flows were carefully controlled through flow meters and, to minimize water vapor, the mixed gases were passed over a desiccant to reduce $p_{H2O}$ to below $10^{-8}$ bar. For annealing experiments in controlled $p_{H2O}$, nitrogen with $10^{-4}$ bar $H_2O$ was mixed into the gas stream at the inlet of the furnace. Gas from the furnace was passed through a glycerin bubbler and vented to atmosphere ensuring no backflow into the furnace. Unless otherwise specified, all samples were annealed under a 1 bar total pressure ($P_{total}$). When not in use, the furnace was continuously purged with 1000 sccm of liquid nitrogen boiloff. All high purity gases were acquired from Airgas and are summarized in Table I. Gas mixtures (1% $O_2$ in $N_2$ and 100 ppm $H_2O$ in $N_2$) do not specify $N_2$ purity. A vacuum port, open only during sample loading, minimized furnace contamination from ambient air. After loading into the furnace, a minimum 20-minute gas purge with the intended high purity ambient was performed prior to moving the sample into the preheated hot zone. A discussion of the importance of purging times is included in the SI. Unless otherwise stated as a second, subsequent, or staged anneal experiment, all results are for as-implanted samples.

TABLE I. High purity gas specifications from AirGas. All impurity levels are given in ppm. The terms R, RP, and UHP are used in the text to designate the specific gases in this table.

| Product Name | Minimum Purity | $O_2$ | $H_2O$ | THC | CO | $CO_2$ | $H_2$ (for $N_2$) | $N_2$ (for Ar) |
|---|---|---|---|---|---|---|---|---|
| Research Plus (RP) $N_2$ | 99.9999% | ≤ 0.2 | ≤ 0.2 | ≤ 0.1 | ≤ 0.3 | ≤ 0.1 | -- | N/A |
| Research (R) $N_2$ | 99.9997% | ≤ 0.5 | ≤ 0.5 | ≤ 0.2 | ≤ 0.5 | ≤ 0.5 | ≤ 2 | N/A |
| Ultra-High Purity (UHP) $N_2$ | 99.999% | ≤ 1 | ≤ 1 | ≤ 0.5 | *≤ 1 | *≤ 1 | -- | N/A |
| RP Ar | 99.9999% | ≤ 0.1 | ≤ 0.2 | ≤ 0.1 | ≤ 0.1 | ≤ 0.1 | N/A | ≤ 2 |
| 100 ppm $H_2O$ in $N_2$ | N/A | N/A | 100 | N/A | N/A | N/A | N/A | N/A |
| 1% $O_2$ in $N_2$ | N/A | 10,000 | N/A | N/A | N/A | N/A | N/A | N/A |

*N/A = not applicable for specified gas*
*-- = not specified*
*\* = CO + $CO_2$ ≤ 1 ppm*

Electrical activation was determined using a Nanometrics HL5500 Hall system with indium contacts made to the corners in a van der Pauw geometry. The active carrier fraction was defined as the ratio of the measured sheet concentration ($n_s$) to the total implant dose (Si/cm$^2$).

Free carrier concentrations within the box implant were then estimated by multiplying the active carrier fraction by the total target implant concentration (Si/cm$^3$). On select samples, contact resistances to the highly doped films were extracted using the transfer length method (TLM). A BCl$_3$/Ar ICP-RIE dry etch (20 W RF and 250 W ICP) with a Ti/Ni hard mask was used for mesa isolation. The Ti/Ni hard mask was stripped using a 1:1 HF:HNO$_3$ solution. A Ti/Al/Ni (50/100/65 nm) stack, deposited by electron-beam evaporation at a base pressure of 4.5×10$^{-7}$ Torr, was patterned via optical lithography and liftoff for metal contacts. TLM samples were annealed in a series of 5 s RTA cycles in N$_2$ from 300 to 480 °C in steps of 30 °C to ensure ohmic contact formation between the metal and the highly doped films.

X-ray diffraction (XRD) was performed using a PANalytical Empyrean diffractometer with Cu K$_{\alpha 1}$ radiation. Rutherford backscattering spectrometry in channeling mode (RBS/c) was performed with a Model 3S-MR10 accelerator from National Electrostatics Corporation (NEC) calibrated using indium zinc oxide (IZO) on glassy carbon. Data were collected for each as-implanted and annealed sample in a 168° backscattering geometry with 2 MeV He$^+$ beam energy, and 40 µC per scan with one scan each in random and channeling configurations. Cross-sectional scanning transmission electron microscopy (STEM) samples were prepared using a Thermo Fisher Helios G4 UX Focused Ion Beam with a final milling step of 5 keV. A carbon layer was deposited to reduce charging during sample preparation. STEM imaging was performed with an aberration corrected Thermo Fisher Spectra 300 CFEG operated at 300 keV.

## III. RESULTS AND DISCUSSION

### Electrical Activation

Previous annealing studies using $N_2$ ambients did not quantify $p_{O2}$ levels, and as noted earlier, some $p_{O2}$ is necessary to stabilize $Ga_2O_3$ against decomposition. To study the $p_{O2}$ dependence of electrical activation, samples implanted to each of the three Si concentrations were annealed under varying $p_{O2}$ using either UHP $O_2$ or 1% $O_2$ in $N_2$ gas mixed with RP $N_2$ ($P_{total}$ = 1 bar, $p_{H2O} < 10^{-8}$ bar for all conditions). After sample loading, the furnace was purged for 20 minutes with 2000 sccm and samples were annealed for 10 minutes at 950 °C. For the lowest $p_{O2}$ ($< 2 \times 10^{-7}$ bar, RP $N_2$), measured sheet resistances ($R_s$) were 1260, 161, and 199 Ω/□ for $5 \times 10^{18}$, $5 \times 10^{19}$, and $1 \times 10^{20}$ cm$^{-3}$ samples, respectively. $R_s$ for the $1 \times 10^{20}$ cm$^{-3}$ sample was higher than for the $5 \times 10^{19}$ sample as 10 minutes at 950 °C is beyond the optimal time for $1 \times 10^{20}$ cm$^{-3}$ implants, as discussed later. Figure 2 shows the relative activation as a function of $p_{O2}$, with the relative sheet resistance ($R_{s,rel}$) defined as $R_s$ divided by $R_s$ at $p_{O2} < 2 \times 10^{-7}$ bar ($R_{s,low\ PO2}$); shaded regions correspond to $R_{s,rel.} > 2$. For $5 \times 10^{18}$ cm$^{-3}$, implants activated for anneals over the full range of $2 \times 10^{-7}$ bar $< p_{O2} < 1$ bar with no significant change in $R_s$, mobility, or activation fraction. For $5 \times 10^{19}$ cm$^{-3}$ implants, activation was insensitive to $p_{O2}$ up to $10^{-4}$ bar, but at $10^{-3}$ bar, $R_s$ increased by several orders of magnitude. At $1 \times 10^{20}$ cm$^{-3}$, carriers activated only under RP $N_2$, $p_{O2} < 2 \times 10^{-7}$ bar; any additional $O_2$ caused $R_s$ to increase dramatically (an order of magnitude at $10^{-5}$ bar $p_{O2}$). The decreased tolerance to oxygen at high doping concentrations is likely due to the increased $p_{O2}$ and Fermi level causing an increase in $V_{Ga}$ concentrations.[7]

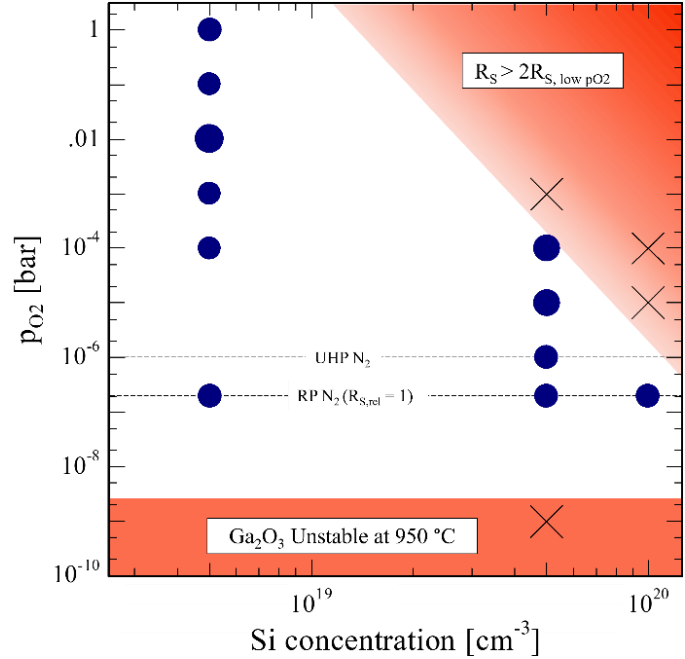

FIG 2. $p_{O2}$ dependence of Si activation at 950 °C for 10 minutes as a function of implant concentration. Experimental conditions for $R_s$ values below $2R_{s,low\ pO2}$ ($p_{O2} < 2\times10^{-7}$ bar) are shown as filled circles with the diameter of the circle representing the $R_{s,rel}$ value (with $R_{s,low\ pO2}$ = "1"); black "×" symbols indicate conditions that do not result in high activation ($R_{s,rel} > 2$). Regions shaded in red are guides to the eye and represent $p_{O2}$ values with $R_{s,rel} > 2$.

Activation behavior was equivalent during anneals in argon ($p_{O2} < 2\times10^{-7}$ bar), confirming that $N_2$ is not critical. The lower bound on $p_{O2}$, set by $\beta$-$Ga_2O_3$ stability, was evident in anneals under UHV ($P_{total} < 2\times10^{-9}$ bar and $pO2 < 10^{-9}$ bar) and forming gas (4% $H_2$ in $N_2$, $pO2 < 10^{-12}$ bar). UHV anneals of $5\times10^{19}$ cm$^{-3}$ resulted in $R_s$ of 1-2 k Ω/□ (tenfold increase, indicated as black "×" in Figure 2), and anneals under forming gas decomposed the $Ga_2O_3$ (Fig. S2). Under UHV, it is unlikely that the effect is a result of the change in total pressure but rather a result of the decreased $p_{O2}$. This establishes a lower bound for annealing at 950 °C of $p_{O2} > 10^{-9}$ bar. While lower Si concentration implants tolerate $p_{O2}$ up to 1 bar, there does not appear to be any advantage to higher $p_{O2}$ for activation.

While activation was achieved over a wide range of $p_{O2}$, anneals were much more sensitive to trace $p_{H2O}$ contamination. As-implanted samples at $5\times10^{19}$ cm$^{-3}$ were annealed for 20 minutes at 950 °C under R N$_2$ (P$_{total}$ = 1 bar, $p_{H2O}$ < $1\times10^{-8}$ bar, $p_{O2}$ < $5\times10^{-7}$ bar) mixed with controlled amounts of H$_2$O; $p_{H2O}$ values of < $1\times10^{-8}$, $2.5\times10^{-7}$, $2.5\times10^{-6}$, and $2.5\times10^{-5}$ bar were tested. Figure 3 shows $R_s$, mobility, and percent carrier activation as a function of $p_{H2O}$ after the initial anneal (blue). Even at $2.5\times10^{-7}$ bar $p_{H2O}$ (0.25 ppm), the activation fraction decreased. By $2.5\times10^{-5}$ bar $p_{H2O}$, $R_s$ increased by an order of magnitude as $n_s$ decreased tenfold. As Figure 3 indicates, mobility decreased only slightly with the initial H$_2$O addition but then increased at high $p_{H2O}$ (likely due to reduced scattering with the lower $n_s$). Presence of H$_2$O became more detrimental with the addition of O$_2$ to the ambient, as discussed in the SI. Subsequent annealing for 20 minutes (orange) in dry, low $p_{O2}$ nitrogen at 950 °C showed partial recovery of properties. For $p_{H2O}$ < $10^{-6}$ bar (<1 ppm), the additional 20-minute dry anneal resulted in "over-annealing" (discussed below) and a slight increase in $R_s$. These data show that the impact of annealing in a wet ambient is largely recoverable, but for high implant activation $p_{H2O}$ must be held to <$10^{-8}$ bar. Reducing $p_{H2O}$ in the system to this level requires an extended gas purge before annealing, as discussed further in the SI.

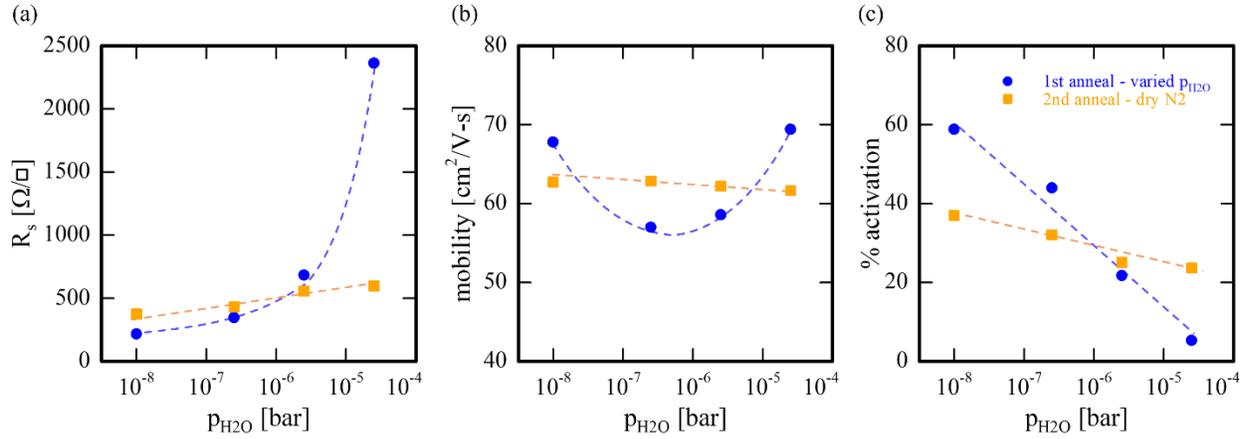

FIG 3. Plots of $R_s$ (a), $\mu$ (b), and % activation (c), as a function of $p_{H2O}$ added to otherwise high purity (R, dry <10 ppb $H_2O$) $N_2$, showing first anneal (950 °C, 20 minutes) with varied $H_2O$ content in blue and after a second anneal (950 °C, 20 minutes) under dry $N_2$ in orange, showing recovery of electrical properties. Dashed lines are added as a guide to the eye.

Based on the findings of the impact of $p_{O2}$ and $p_{H2O}$, the annealing behavior with time and temperature was measured under $N_2$ ($p_{O2} < 10^{-6}$, $p_{H2O} < 10^{-8}$ bar, $P_{total} = 1$ bar) for $5\times10^{19}$ cm$^{-3}$ implants over 2.5 to 30 minutes at temperatures from 850 to 1050 °C, using sequential anneals of single samples to minimize sample variation errors. After sample loading, the furnace was purged for 20 minutes. Following each anneal, indium contacts were soldered to the corners of the samples, Hall measurements were obtained, and the indium contacts were stripped with HCl; the sample was then loaded for the next anneal step. It is important to note that the "total anneal time" does not correct for the finite time required to reach the set temperature after transfer into the furnace (approximately 2 minutes). However, the staged annealing does provide monotonic trends with time. Figure 4 shows these staged time annealing results for temperatures from 850 to 1050 °C. The % activation (Fig. 4c) is the percent of the measured sheet concentration, $n_s$, compared to the total implanted dose. At lower temperatures (850 to 900 °C) there was a strong annealing time dependence to $R_s$ (Fig. 4a), which decreased for times up to 30 minutes as the mobility (Fig. 4b) and carrier activation (Fig. 4c) increased. Even after 30 minutes, the mobility did not reach the level observed for higher temperature anneals. At higher temperatures (1000 and 1050 °C), the mobility saturated at the shortest anneals indicating that the implant damage rapidly recovered. However, Si is known to diffuse at temperatures >950 °C [23,34] limiting useful annealing to lower temperatures. With extended time at higher temperatures, $R_s$ began to increase associated with a decrease in carrier activation, an "over-annealing" behavior discussed further below. With these considerations, 950 °C emerges as an optimized annealing temperature with a broad anneal time window of 5-30 minutes; $R_s$ reached a minimum after only 5 minutes (corresponding with maximized µ and % activation) and held for 30-40 minutes before $R_s$ started to slowly increase.

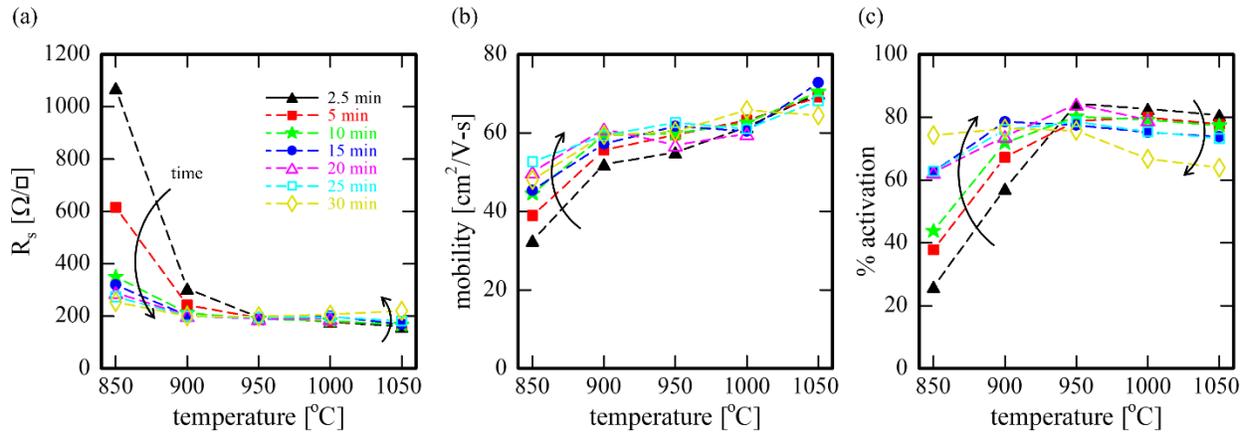

FIG 4. Plots of $R_s$ (a), µ (b) and % activation (c) versus anneal temperature for different times (indicated by different colors and symbols in the legend), showing trends with time and temperature for anneals under dry $N_2$. Arrows indicate trends with increasing time, showing a decrease in $R_s$ (a), increase in µ (b), and an increase in carrier activation (c) with time at lower temperatures and a slight increase in $R_s$ (a) with a decrease in carrier activation (c) with time at higher temperatures.

For long anneal times, the activation fraction decreased resulting in "over-annealing". At $5\times10^{19}$ cm$^{-3}$, deactivation was observed even at low temperatures, occurring after 40 minutes at 950 °C but after only 15 minutes at 1050 °C. Over-annealing in implanted samples manifested as a decrease in active carriers at all anneal temperatures, and as a decrease in mobility at the highest temperatures. To investigate the dose dependence, samples at $5\times10^{18}$ and $1\times10^{20}$ cm$^{-3}$ were also time-stage annealed with Hall measurements after each step. Figure 5 shows the properties as a function of anneal time (950 °C, $p_{O2} < 10^{-6}$, $p_{H2O} < 10^{-8}$ bar, $P_{total} = 1$ bar). For $5\times10^{18}$ cm$^{-3}$ implants, there was minimal change in activation up to 60 minutes with >70% activation and μ >90 cm$^2$/Vs. For $5\times10^{19}$ cm$^{-3}$, anneals for up to 20 minutes showed 80% activation with μ~60 cm$^2$/Vs, followed by a decrease in carrier activation and a rise in $R_s$ at longer times. At the highest implant concentration, $1\times10^{20}$ cm$^{-3}$, there was a very strong time dependence with significant over-annealing even after 10 minutes. For a 5 minute anneal, 81% of the carriers were activated (estimated concentration of $8.14\times10^{19}$ cm$^{-3}$) with a mobility of 70.8 cm$^2$/Vs and $R_s$ of 75.3 Ω/□. All subsequent anneals reduced the carrier activation and after 60 minutes $R_s$ increased threefold to 228 Ω/□ with only 28% of the implanted carriers activated. Earlier reports in literature have suggested that elevated temperature during implantation is required to activate $1\times10^{20}$ cm$^{-3}$ Si concentration.[16] Our results, however, show that $1\times10^{20}$ cm$^{-3}$ implants can be almost fully activated with high mobility if $p_{O2}$, $p_{H2O}$, and time are carefully controlled. As the over-annealing is highly correlated to Si concentration, it is likely related to the formation of defect pairs[7] or clustering of $Si_{Ga}$ substitutions as seen after annealing at elevated temperatures.[35]

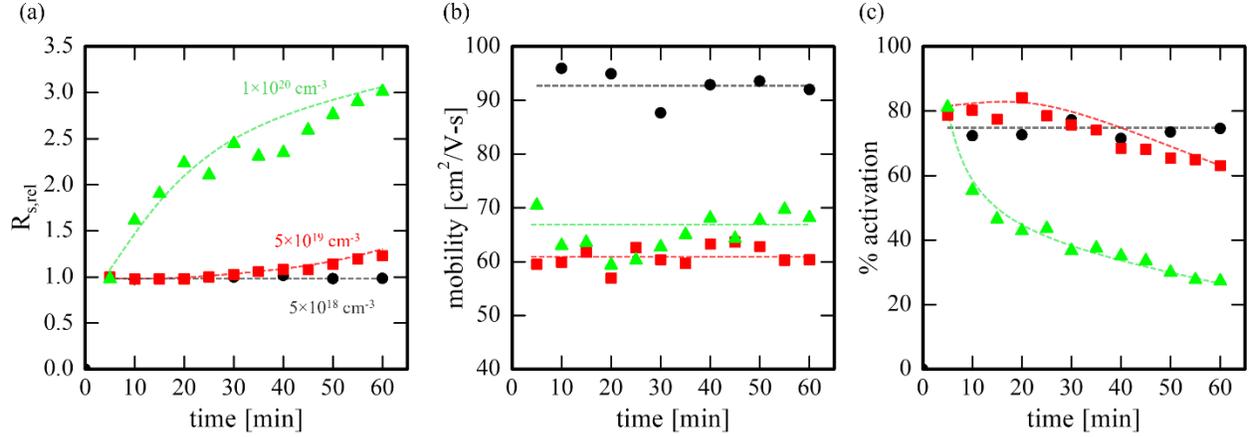

FIG 5. Plots of $R_{s,rel.}$ (= $R_s$ / $R_{s,t=5min}$) (a), $\mu$ (b) and % activation (c) versus time for implant conditions $5\times10^{18}$ cm$^{-3}$ (black), $5\times10^{19}$ cm$^{-3}$ (red), and $1\times10^{20}$ cm$^{-3}$ (green) for anneals under dry $N_2$. No evidence of over-annealing is seen in the $5\times10^{18}$ cm$^{-3}$ sample, minimal over-annealing in the $5\times10^{19}$ cm$^{-3}$ sample after 20 minutes, and significant over-annealing in the $1\times10^{20}$ cm$^{-3}$ sample after the initial 5 minute anneal. Lines are added as a guide to the eye.

To determine if over-annealing is a result of implant-induced damage, an *in-situ* doped $6.9\times10^{19}$ cm$^{-3}$ MOCVD sample was subjected to the same staged anneals at 950 °C. As for implanted samples, $R_s$ increased from 40 (as grown) to 190 Ω/□ after 30 minutes with the mobility decreasing from 91 to 58 cm$^2$/V-s and carriers deactivating to $2.3\times10^{19}$ cm$^{-3}$. These results suggest that the deactivation is not primarily a result of implant damage, but that over-annealing is associated with extrinsic defects in the film and substrate. Mechanisms may differ between implanted and *in-situ* doped samples with the deactivation of the *in-situ* doped samples also involving a decrease in mobility even at 950 °C.

Figure 6 compares deactivation as a function of staged time for samples implanted at $5\times10^{19}$ cm$^{-3}$ from two different epitaxial growth runs. Results show that the rate of deactivation is dependent on the sample position within the 23×25 mm wafers (likely substrate variations) and is potentially dependent on precise conditions during the epitaxial growth. Samples from the two growths were annealed together to ensure identical thermal histories, with one pair annealed in 10-minute increments and the second pair in 20-minute increments. Both pairs show differences in deactivation rates, but the slower deactivating sample in each run came from different epitaxial growth runs. These results indicate growth induced defects, which may vary by substrate position and growth method, are likely important in determining the rate of deactivation. The specific defect contributing to this effect is not known but may be related to the density of screw dislocations[23] facilitating the formation of deactivated defect pairs.

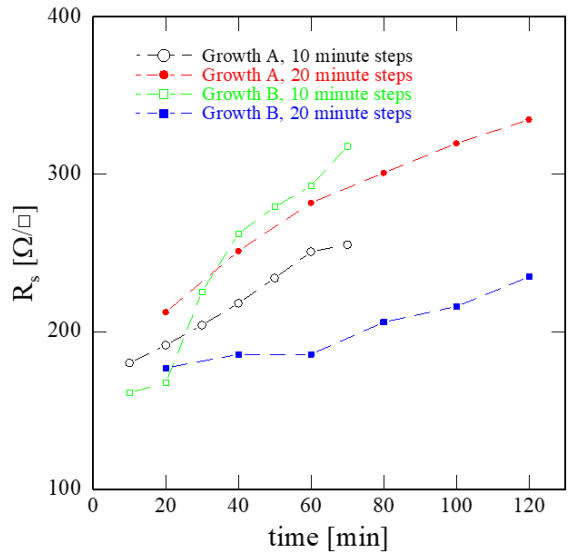

FIG 6. (a) Plots of $R_s$ vs time for $5\times10^{19}$ cm$^{-3}$ implanted samples from two PAMBE growths (A and B) and two anneal time steps (10 and 20 minutes) showing that over-annealing begins after 20 minutes (at 950 °C, $p_{O2} < 10^{-6}$, $p_{H2O} < 10^{-8}$ bar, $P_{total} = 1$ bar), but the rate varies depending on specific growth runs and position on the substrate.

The optimized anneal conditions for activating Si in $\beta$-Ga$_2$O$_3$ vary with implant concentrations, becoming more restrictive as the Si concentration increases. The annealing ambient is critical for activation, requiring $p_{H2O} < 10^{-8}$ bar and $10^{-9} < p_{O2} < 10^{-6}$ bar. Annealing at 950 °C for 5-20 minutes is sufficient to fully recover the lattice and mobility while minimizing over-annealing, even for concentrations to $1\times10^{20}$ cm$^{-3}$. Figure 7 compares these Si mobility and carrier activation results from implant and annealing with other literature results for Si in [010] $\beta$-Ga$_2$O$_3$, as well as the MOCVD *in-situ* doped sample results discussed in this paper. For the three implants studied in this work, we observed 94.0 cm$^2$/Vs at $3.95\times10^{18}$ cm$^{-3}$, 71.6 cm$^2$/Vs at $4.22\times10^{19}$ cm$^{-3}$, and 70.8 cm$^2$/Vs at $8.15\times10^{19}$ cm$^{-3}$. These results confirm that implant and thermal anneals are competitive with *in-situ* doping methods. The ability to form good ohmic contacts with an average contact resistance $R_c = 0.29 \pm 0.02$ Ω-mm was demonstrated by TLM measurements in the $5\times10^{19}$ cm$^{-3}$ samples annealed at 950 °C (see SI).

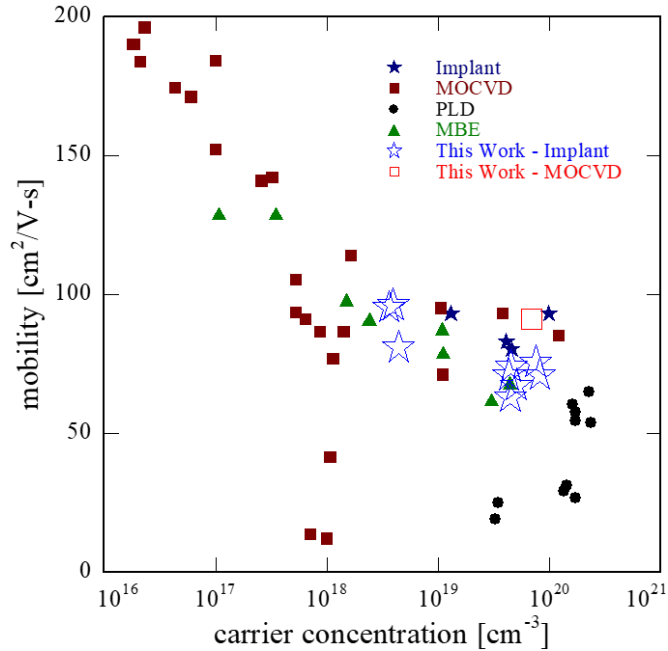

FIG 7. Blue open stars represent mobility and carrier concentration values for implant samples reported in this study (red open square for *in-situ* MOCVD) compared with select literature reports for [010] $\beta$-Ga$_2$O$_3$ samples doped with Si by implant,[16,18] MOCVD,[9,13,36–38] PLD,[39–41] and MBE.[11,12]

Implant Damage and Lattice Recovery

To investigate the damage and subsequent recovery, XRD, RBS/c, and STEM were used to analyze samples implanted to $5\times10^{19}$ cm$^{-3}$. Figure 8a shows XRD scans for pre-implant (gray), post-implant (blue), and post-anneal (orange) for 20 minutes at 950 °C. No additional peaks were seen in the full range 2θ scans after implant or after annealing, though the implanted sample does exhibit shoulders around the (020) peak indicating strain within the film; this damage is recovered with annealing. Phase transformations that may have been induced by the implant were not detected by XRD for implant concentrations up to $1\times10^{20}$ cm$^{-3}$. Additional XRD scans, including rocking curves, are included in the SI.

Damage accumulation and recovery was further investigated with RBS/c, as shown in Figure 8b for an as-implanted and annealed (950 °C, 20 minutes, dry N$_2$) $5\times10^{19}$ cm$^{-3}$ sample. The random spectrum, with slight planar channeling, matches well with RBS simulations[42,43] of pure Ga$_2$O$_3$ (blue). Channeling of the as-implanted sample (red) indicates only partial damage with no fully amorphous layer; the maximum scattering is only 70% of the expected fully amorphous level (scattering expected for a 200 nm Ga$_2$O$_3$ film layer is shown in magenta). After annealing, channeling (green) shows full recovery of the crystal lattice with a $\chi_{min}$ of 2.8%, consistent with a good single crystal; the de-channeling with depth is also typical for (010) oriented films.[35]

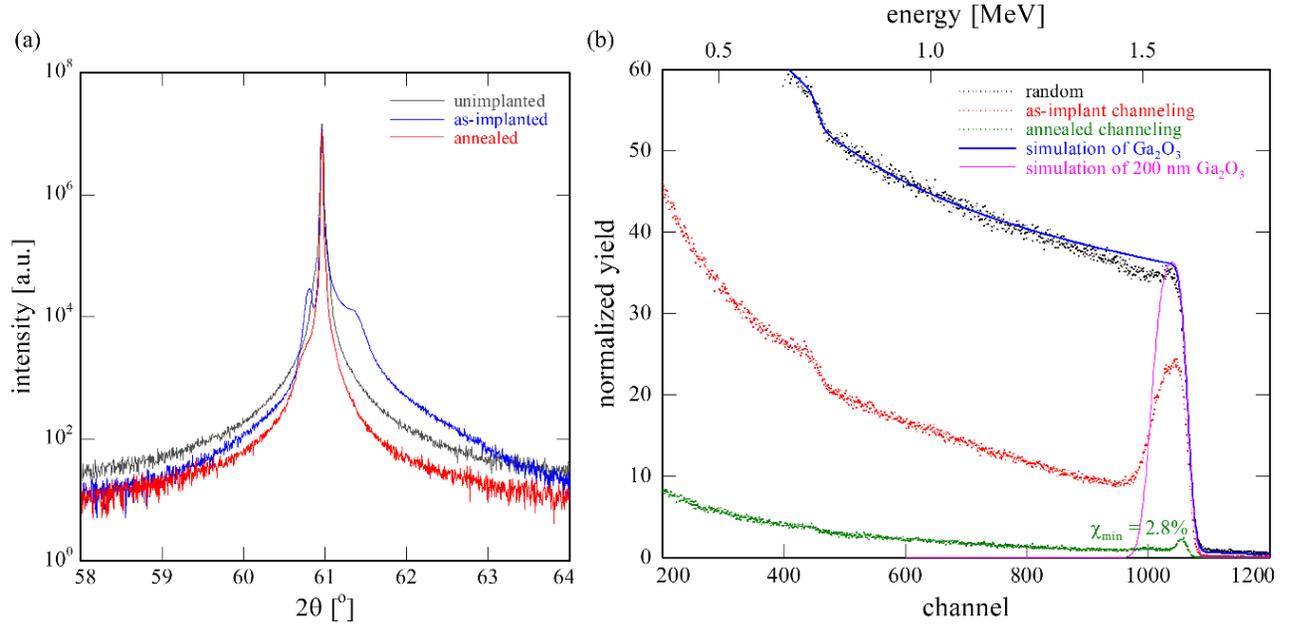

FIG 8. (a) θ-2θ XRD patterns of (020) reflection for un-implanted (gray), as-implanted to $5\times10^{19}$ cm$^{-3}$ (blue, showing shoulders associated with lattice damage), and implanted and annealed (950 °C, 20 minutes, N$_2$, red); and (b) RBS/c spectra compared to simulated spectra showing random (black and blue line, respectively); implanted to $5\times10^{19}$ cm$^{-3}$ channeling in red, showing peak yield in implanted region not reaching full random signal but matching projected depth of damage (simulated 200 nm in magenta); and after annealing (20 minutes, 950 °C dry N$_2$), recovery of crystal lattice to near perfect structure in green with $\chi_{min}$ of 2.8%.

Figure 9 summarizes high-angle annular dark field (HAADF)-STEM measurements performed on a sample implanted to $5\times10^{19}$ cm$^{-3}$ Si. The image shows defects identified as Ga$_i$ interstitials (blue arrows), regions of retained *β*-phase (green), and regions of both [110] *γ*-phase and overlapping *γ*-phase (pink), as has been repeatedly observed in literature as implant-induced phase transformations.[21–23] This observed phase transformation supports previous findings that *γ*-Ga$_2$O$_3$ is the kinetically favored structure, often forming in regions of high disorder such as substrate interfaces, free surfaces, and areas with high implant damage.[44] To highlight the implant damage, the Fast Fourier Transform (FFT) of the damaged lattice (Fig. 9a) is shown in Figure 9b with the FFT of the annealed, recovered lattice overlaid in green (HAADF-STEM image shown in Figure S8). Overlapping FFT peaks appear white, confirming the presence of retained β-phase in the damaged lattice; additional diffraction spots present only in the damaged FFT are shown in magenta. Figure 9c shows the predicted electron diffraction patterns[45] of both the β-phase along the [010] axis (green) and the γ-phase along the [110] axis (magenta), showing that the overlayed patterns match the FFT of the implanted area, confirming the presence of the γ-phase in the implanted region. Additional images in the SI show a comparable analysis from 90° rotated from the [010] axis.

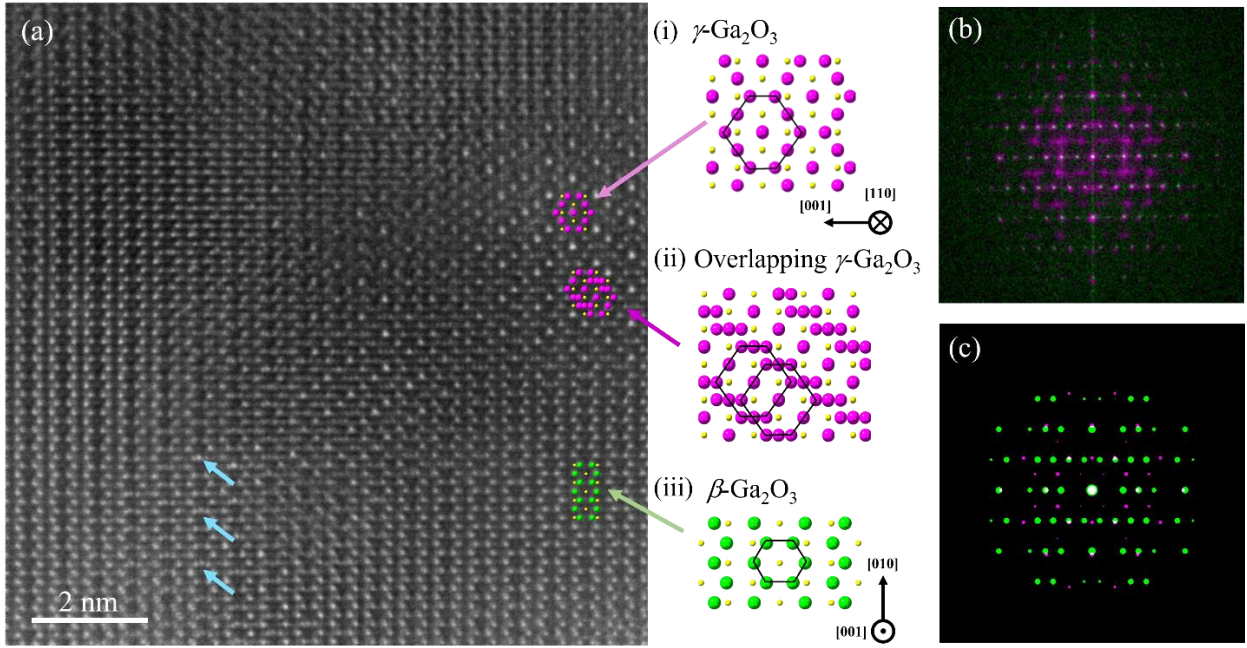

FIG 9. (a) Atomic resolution HAADF-STEM image of as-implanted film, highlighting regions of transformed γ-phase (i) and overlapping γ-phase (ii), $Ga_i$ interstitials (blue arrows), and retained β-crystallinity (iii). Green and magenta represent Ga atoms (in β and γ, respectively) and yellow represents O atoms. (b) FFT of as-implanted region shown in (a) overlaid with the FFT of β-phase crystal. The magenta shows the additional damage from implant and white represents areas with intensity from both FFT patterns. Note no pure green spots are observed in the FFT as the implanted region contains both β– and γ-phases. (c) Simulated single crystal electron diffraction

patterns along the [010] zone axis of β-Ga$_2$O$_3$ and [110] of γ-Ga$_2$O$_3$, showing correlation with the measured pattern in (b).

The combined structural information from XRD, RBS/c, and STEM confirms remnant crystallinity in the as-implanted films for the 5×10$^{19}$ cm$^{-3}$ Si implant. It is hypothesized that this remnant crystallinity seeds recovery of the lattice upon annealing, without requiring epitaxial regrowth from below the depth of damage or the substrate.

## IV. CONCLUSIONS

With careful control of annealing $p_{O2}$, $p_{H2O}$, temperature and time, Si implant concentrations in β-Ga$_2$O$_3$ from 5×10$^{18}$ to 1×10$^{20}$ cm$^{-3}$ can be highly activated (>80%) with full recovery of the mobility to >70 cm$^2$/Vs. In the ambient, $p_{O2}$ must be above 10$^{-9}$ bar to maintain stability of the Ga$_2$O$_3$, with the upper bound dependent on Si concentration (>1 bar for 5×10$^{18}$ cm$^{-3}$ and 10$^{-6}$ bar for 1×10$^{20}$ cm$^{-3}$). Water must be minimized during implant annealing with $p_{H2O}$ < 1×10$^{-8}$ bar as even 2.5×10$^{-7}$ bar $p_{H2O}$ reduces active carriers and increases R$_s$; the impact is even stronger when O$_2$ is also present in the gas ambient ($p_{O2}$ > 10$^{-6}$ bar). 950 °C is an optimal temperature for activation of all three implant concentrations, maximizing the recovered mobility with minimal diffusion. Anneal time is also critical, especially for high concentrations, with 5 minutes in a traditional furnace sufficient to activate implants at 950 °C. The upper time limit for annealing is set by the onset of deactivation and depends on Si concentration. At low Si concentrations, deactivation is not observed after even 60-minute anneals while high Si concentrations begin to deactivate within 10 minutes. Investigations into the lattice damage and recovery show a high degree of retained β-Ga$_2$O$_3$ crystallinity in as-implanted regions which rapidly seeds lattice recovery and enables annealing at the relatively low temperature of 950 °C.

For $5\times10^{19}$ cm$^{-3}$ implants, contact resistances below 0.29 Ω-mm can readily be achieved, showing promise for selective area doping methods. Mobility as a function of carrier concentration for implants is comparable to the best reports from *in-situ* doped methods showing that ion implantation is a highly competitive doping method in $\beta$-Ga$_2$O$_3$.

## SUPPLEMENTAL INFORMATION

See the supplemental material for additional implant details, a furnace schematic, a discussion of the importance of purge times, results of annealing Ga$_2$O$_3$ in forming gas, contact resistance analysis, additional XRD plots, and additional STEM images.

## ACKNOWLEDGMENTS

This research was supported by the Air Force Research Laboratory-Cornell Center for Epitaxial Solutions (ACCESS) under Grant No. FA9550-18-1-0529. The authors also acknowledge the use of the Cornell Center for Materials Research Shared Facilities supported through the NSF MRSEC program (DMR-1719875) and the Cornell NanoScale Facility, a member of the National Nanotechnology Coordinated Infrastructure (NNCI), which is supported by the National Science Foundation (Grant No. NNCI-2025233).

## AUTHOR DECLARATIONS

### Conflicts of Interest

The authors have no conflicts to disclose.

## DATA AVAILABILITY

The data that support the findings of this study are available from the corresponding author upon reasonable request.

# Supplemental Information for "Silicon Implantation and Annealing in $\beta$-Ga$_2$O$_3$: Role of Ambient, Temperature, and Time"

**Sample Summaries:**

Table SI shows relevant details for separate growths and implants including substrate ID (from NCT) and the specific implant parameters. The key experiments presented in the paper are listed in Table SII and specify the particular implanted sample(s) used. Due to inherent variation between substrates as well as epi growths, each set of experiments conducted utilized samples from the same growth.

TABLE SI. Table of implant energies and doses for each growth and substrate ID as well as the experimental results obtained from each sample.

| Implant #: | NCT Sub. ID: | Box Concentration: | Total Dose: | Implant Sequence: |
|---|---|---|---|---|
| 1 | N00221 | $5 \times 10^{19}$ cm$^{-3}$ | $7.11 \times 10^{14}$ cm$^{-2}$ | $6.1 \times 10^{13}$ cm$^{-2}$ @ 16 keV<br>$1.4 \times 10^{14}$ cm$^{-2}$ @ 46 keV<br>$5.1 \times 10^{14}$ cm$^{-2}$ @ 120 keV |
| 2 | N00224 | $5 \times 10^{19}$ cm$^{-3}$ | $7.16 \times 10^{14}$ cm$^{-2}$ | $5.60 \times 10^{13}$ cm$^{-2}$ @ 15 keV<br>$1.40 \times 10^{14}$ cm$^{-2}$ @ 45 keV<br>$5.20 \times 10^{14}$ cm$^{-2}$ @ 115 keV |
| 3 | N01692 | $5 \times 10^{19}$ cm$^{-3}$ | $6.88 \times 10^{14}$ cm$^{-2}$ | $4.30 \times 10^{13}$ cm$^{-2}$ @ 12 keV<br>$1.35 \times 10^{14}$ cm$^{-2}$ @ 40 keV<br>$5.10 \times 10^{14}$ cm$^{-2}$ @ 112 keV |
| 4 | N01683 | $5 \times 10^{18}$ cm$^{-3}$ | $7.16 \times 10^{13}$ cm$^{-2}$ | $5.60 \times 10^{12}$ cm$^{-2}$ @ 15 keV<br>$1.40 \times 10^{13}$ cm$^{-2}$ @ 45 keV<br>$5.20 \times 10^{13}$ cm$^{-2}$ @ 115 keV |
| 5 | N01683 | $1 \times 10^{20}$ cm$^{-3}$ | $1.43 \times 10^{15}$ cm$^{-2}$ | $1.12 \times 10^{14}$ cm$^{-2}$ @ 15 keV<br>$2.80 \times 10^{14}$ cm$^{-2}$ @ 45 keV<br>$1.04 \times 10^{15}$ cm$^{-2}$ @ 115 keV |

TABLE SII. Table of experiments with corresponding figures in the main paper with the implant number from Table SI.

| Experiment: | Figure: | Implant #: |
|---|---|---|
| SIMS comparison to SRIM | 1 | 1 |
| $p_{O2}$ dependence of $5 \times 10^{18}$, $5 \times 10^{19}$, $1 \times 10^{20}$ cm$^{-3}$ at 950 °C | 2 | 3,4,5 |
| $p_{H2O}$ dependence of $5 \times 10^{19}$ cm$^{-3}$ at 950 °C | 3 | 2 |
| Time and temperature dependence of $5 \times 10^{19}$ cm$^{-3}$ | 4 | 3 |
| Over-annealing at 950 °C of $5 \times 10^{18}$, $5 \times 10^{19}$, $1 \times 10^{20}$ cm$^{-3}$ | 5 | 3,4,5 |
| Over-annealing variation of $5 \times 10^{19}$ cm$^{-3}$ at 950 °C | 6 | 2,3 |
| Mobility vs concentration literature comparison | 7 | 2,3,4,5 |
| XRD and RBS/c $5 \times 10^{19}$ cm$^{-3}$ implant | 8 | 2 |
| STEM analysis $5 \times 10^{19}$ cm$^{-3}$ implant | 9 | 2 |
| Vacuum annealing | No figure | 1,2 |
| RP Argon Annealing | No figure | 2 |
| Comparison of UHP, R, RP N$_2$ for $5 \times 10^{19}$ cm$^{-3}$ | No figure | 1,2,3 |

A schematic of the furnace setup is shown in Figure S1. Four flowmeters, with ranges of 5-54, 30-320, or 200-2400 sccm, enabled mixing of gases through either the wet or dry mixing line. The "wet" line contained, for some experiments, intentional $p_{H2O}$ levels; in contrast, the "dry" line was never intentionally exposed to H$_2$O. Additionally, the dry line was optionally passed over a desiccant to further reduce water to < 10 ppb ($10^{-8}$ bar). Dry and wet lines then mix at the inlet to the furnace. Samples were loaded through a vacuum compatible load lock onto a quartz boat, which was then pushed into the hot zone after a sufficient gas purge time (typically 20 minutes at 2000 sccm). Gas effluent from the furnace flowed through a glycerin bubbler to ensure no backflow of ambient air. When not in use, 1000 sccm of liquid nitrogen boiloff nitrogen flowed continuously through the furnace held at 250 °C and $P_{total}$ = 1 bar.

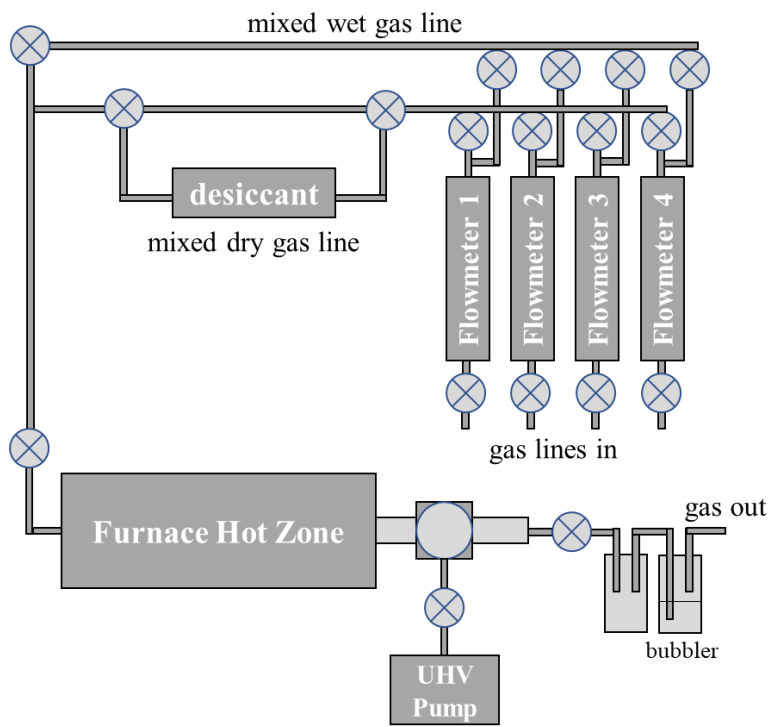

FIG S1. Schematic representation of furnace setup.

**Importance of Purge Times:**

Residual $H_2O$ and $O_2$ can dramatically impact the electrical activation of Si implants in β-$Ga_2O_3$. Control of the annealing ambient required not only high purity sources, but also mitigation of residual gases from the laboratory ambient through purging. Table SIII details a set of 6 samples from the same growth and implant series (Implant #1 from Table S1).

TABLE SIII. $R_s$ values for 6 samples with two $p_{O2}$ levels and varied drying steps, including running the gas line over a desiccant to reduce $H_2O$ < 10 ppb and the recorded purge time after loading the sample into the furnace and before pushing the sample into the hot zone.

| $p_{O2}$ < $10^{-6}$ bar | | | $p_{O2}$ ~ $10^{-4}$ bar | | |
|---|---|---|---|---|---|
| Dry Step | Purge Time | $R_s$ (Ω/□) | Dry Step | Purge Time | $R_s$ (Ω/□) |
| no | 5 min | 249 | no | 9 min | 32,300 |
| yes | 5 min | 165 | yes | 5 min | 397 |
| yes | 12 min | 146 | yes | 10 min | 148 |

Two ambients were investigated: $<10^{-6}$ bar $p_{O2}$ (UHP $N_2$) and $10^{-4}$ bar $p_{O2}$ (UHP $N_2$ mixed with 1% $O_2$ in $N_2$) and three purge conditions were investigated. First, a "no dry step" anneal, where the gas was not passed over the desiccant (to reduce $p_{H2O}$ <10 ppb), shows the highest $R_s$ for both ambient conditions: 249 Ω/□ for the lower $p_{O2}$ and 32 kΩ/□ for $p_{O2} \sim 10^{-4}$. The presence of $H_2O$ and $O_2$ in the ambient is significantly more detrimental than just $H_2O$ as evidenced by the 100x increase in $R_s$ for no drying step with $p_{O2} \sim 10^{-4}$ bar. Flowing the gases over the desiccant with a 5-minute purge reduced the $R_s$ for both ambient to 165 and 397 Ω/□, respectively, confirming the impact of even contaminant levels of $H_2O$ as discussed in the main text. Increasing the purge time to 10 minutes results in comparable activation between the two ambient conditions to below 150 Ω/□; in the absence of $p_{H2O}$ and adequate purging, annealing is stable to moderate $p_{O2}$ for $5 \times 10^{19}$ cm$^{-3}$ concentrations. These results are shown in Figure S2 graphically. To ensure removal of water vapor after sample loading, a purge time of >20 minutes was used for all subsequent anneals.

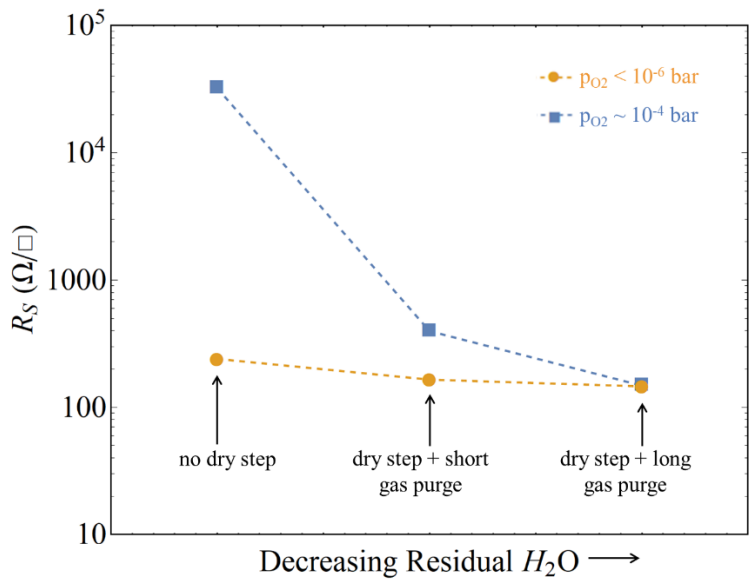

FIG S2. Sheet resistances from Table S3 as a function of the estimated $p_{H2O}$ for annealing with $p_{O2} < 10^{-6}$ and $p_{O2} \sim 10^{-4}$ bar.

**Annealing Under Forming Gas:**

After annealing a sample at 950 °C for 20 minutes under forming gas (4% $H_2/N_2$), the sample was roughened and opaque (left) and condensed, (black) gallium coated the inside of the furnace tube immediately outside of the hot zone.

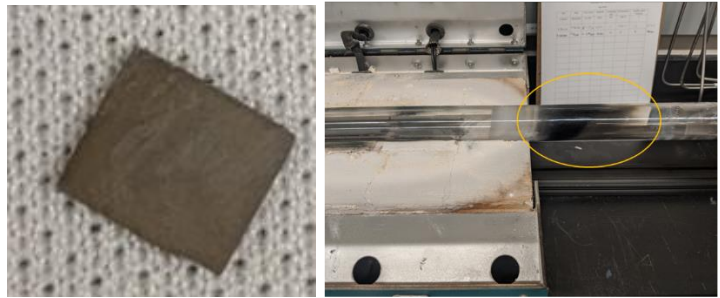

FIG S3. Image of $Ga_2O_3$ sample annealed at 950 °C for 20 minutes under forming gas, showing decomposition of material (left) and image of quartz tube furnace with circled area showing condensation of decomposed Ga immediately outside of the hot zone (right).

**Contact Resistance Measurements:**

The contact resistance was extracted using linear transfer length method (LTLM) patterns with a pad width of 50 μm and pad spacings from 2-12 μm. Pad spacing was confirmed to within ± 0.2 μm for this device process by scanning electron microscopy. The contacts were ohmic, as shown by the IV curves in Figure S4a. The associated TLM extraction is shown in Figure S4b. The contact resistance ($R_c$) extracted by TLM at 1 mA applied bias was $0.31 \pm 0.01$ Ω-mm, and the extracted sheet resistance ($R_s$) was $239 \pm 2$ Ω/□. The associated contact resistivity ($\rho_c$) was $4.1 \pm 0.2 \times 10^{-6}$ Ω-cm². The average contact and sheet resistances, as measured across the eight fabricated LTLM patterns, are shown in Figure S4c. The contact resistance is constant with applied current bias and fairly consistent across TLM patterns. The pattern-to-pattern and pad-to-pad variation is attributed to physical damage of the contact pads during basic IV measurements for contact anneal testing using blunt probe tips. This resulted in significant scratches and wear on some of the contact pads prior to contact resistance measurement.

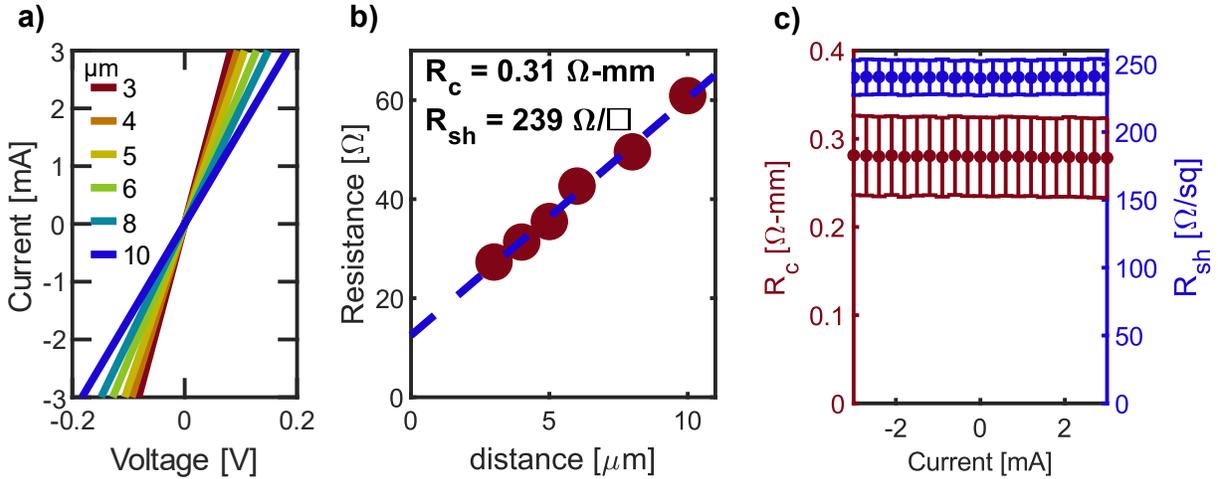

FIG S4. (a) IV curves for a representative LTLM pattern. The contacts are linear ohmic in character, (b) corresponding TLM extraction results in an $R_c$ of 0.31 ± 0.01 Ω-mm and $R_{sh}$ of 239 ± 2 Ω/□, and (c) average $R_c$ and $R_{sh}$ extracted from eight different LTLM patterns are constant with applied current bias and reasonably consistent between patterns.

**XRD Plots:**

Figure S5a shows the (020) rocking curves for samples implanted to $5 \times 10^{19}$ cm$^{-3}$ with as-implanted in blue and after annealing for 20 minutes at 950 °C in orange. The FWHM recovers from 34.7 to 24.2 arcseconds. Figure S5b shows the full range 2θ scans (right), after implant and annealing, showing no additional peaks are observed indicating no substantial volumetric phase transformations during implant.

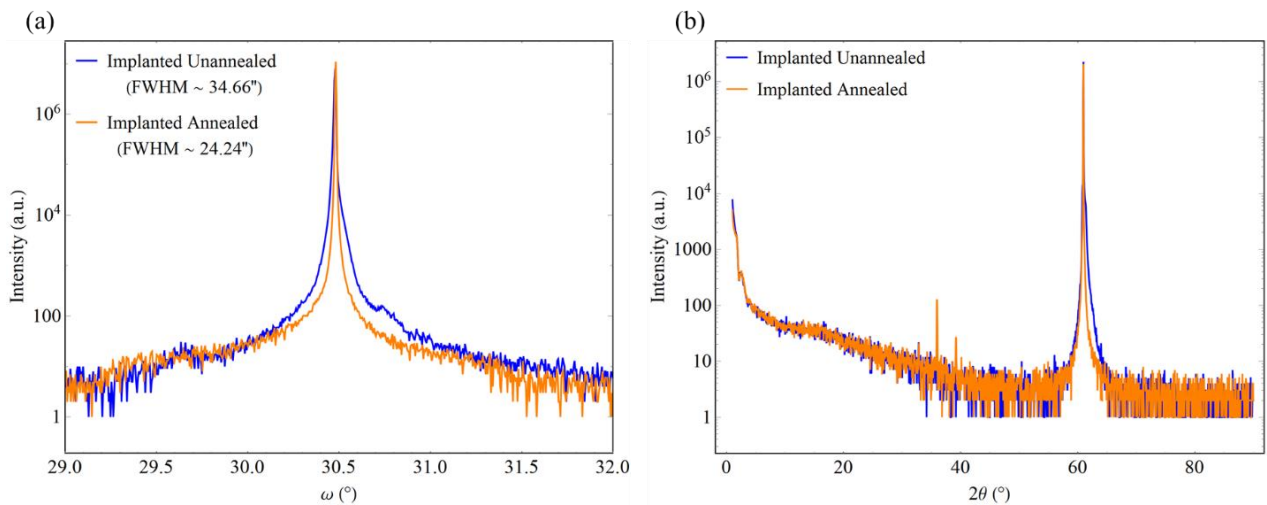

FIG S5. XRD plots of $5 \times 10^{19}$ cm$^{-3}$ implant showing rocking curves (a) and full range $2\theta$ scans (b) for implanted and annealed samples.

Figure S6 shows narrow (a) and full range (b) $2\theta$ XRD scans of a $1 \times 10^{20}$ cm$^{-3}$ sample as-implanted (blue) and after a 950 °C 5 min anneal (orange). Full recovery of the lattice damage is evident in the annealed spectra; the apparent broadening in the full range spectra (b) is due to changes in XRD experimental detector conditions (increased detector channels). The small peak around 60.1° in the spectra on the left is not identifiable as the $\beta$-phase or any other known polymorphs and is likely an instrument artifact. Based on the orientation of the $\gamma$-phase seen in STEM, (110) $\gamma$-phase peaks might be expected, but these were not observed in any implanted samples in this study [(220) at 30.67° or the (440) at 63.87°].

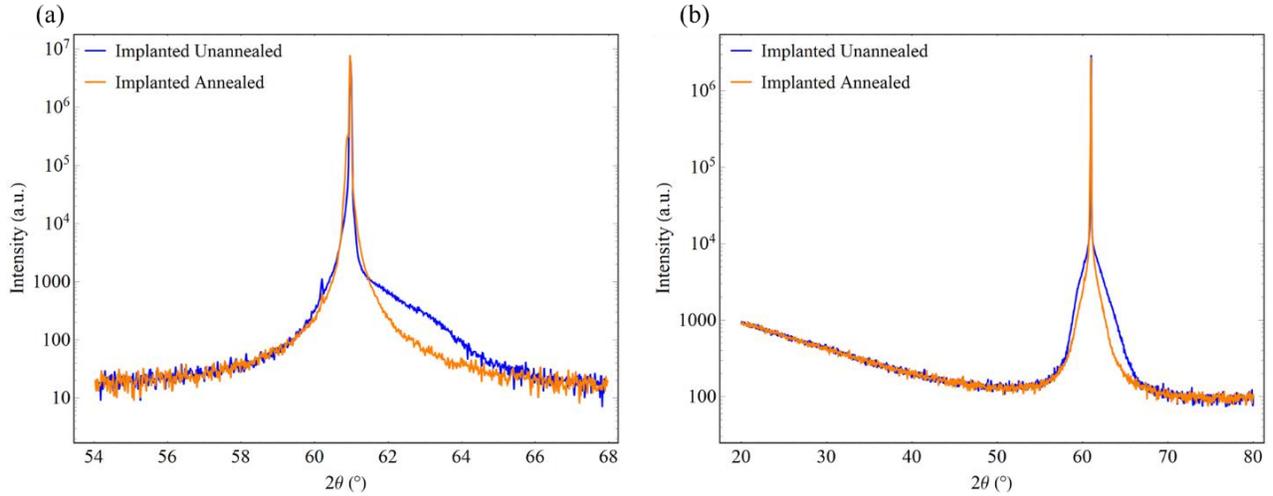

FIG S6. XRD plots for $1\times10^{20}$ cm$^{-3}$ implant samples, showing narrowed range around the (020) Bragg peak (a) and full range 2θ scan (b) showing no secondary phase peaks.

**STEM Images:**

To confirm that the γ-structure observed after implant was not dependent on the imaging direction, a FIB sample was prepared 90° rotated from the first orientation. From the [201] direction (90° rotated from the β[001]), the β-stacking looks very similar to the γ-phase along the [001] direction (90° rotated from the γ[110]), as seen in the simulated structures in Figure S7b and S7c with the similar atomic distances highlighted. While disorder can be seen in Figure S7a, no distinct γ– or β–regions can be identified. The FFT in Figure S7d matches simulated patterns in Figure S6e for mixed γ– and β–phases. For this orientation, many of the γ-peaks nearly overlap the existing β-peaks, but distinct γ-peaks are also present in both the simulated and experimental FFT patterns. While the presence of the γ-phase cannot be directly identified in this imaging orientation, the results are consistent with the direct observation and the orientation of γ– with respect to the β-phase presented in the main text.

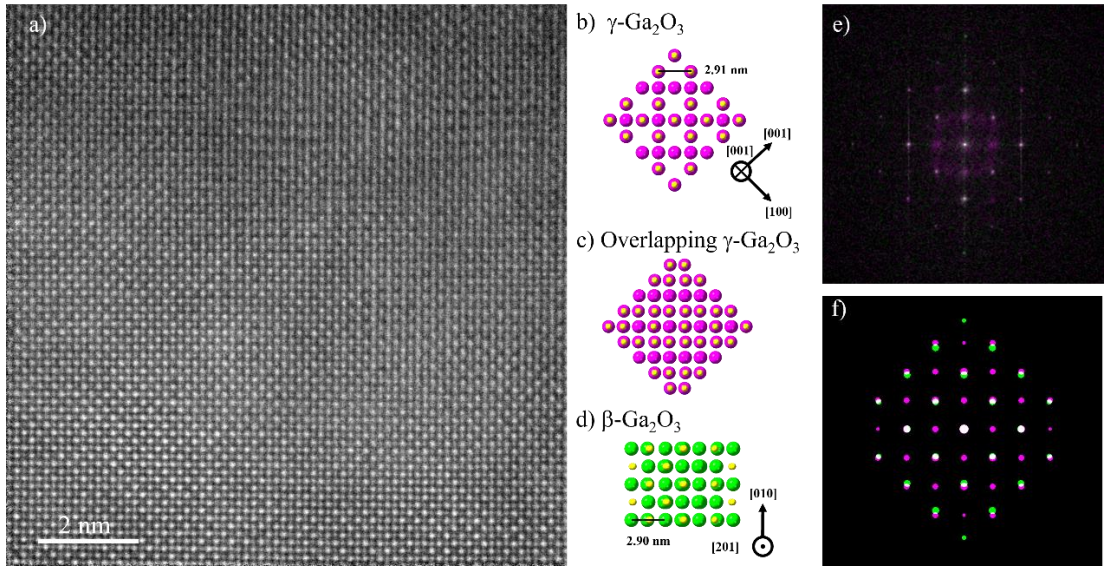

FIG S7. (a) Atomic resolution HAADF-STEM image of implanted film, (b) structure of γ-Ga$_2$O$_3$ along [001] axis, (c) structure of β-Ga$_2$O$_3$ along [$\bar{2}$0$\bar{1}$] axis (d) FFT of implanted region shown in (a) overlaid with the FFT of β-phase crystal. The magenta shows the additional damage from implant and white represents areas with intensity from both FFT patterns. (e) Simulated single crystal electron diffraction patterns along the [$\bar{2}$0$\bar{1}$] zone axis of β-Ga$_2$O$_3$ and [001] of γ-Ga$_2$O$_3$.

Figure S8 shows the recovered β-lattice after annealing a 5x10$^{19}$ cm$^{-3}$ implanted sample at 950 °C for 20 minutes, supporting the lattice recovery seen in RBS/c and XRD. FFT of the recovered lattice shows only β-phase (inset in Figure S8b.

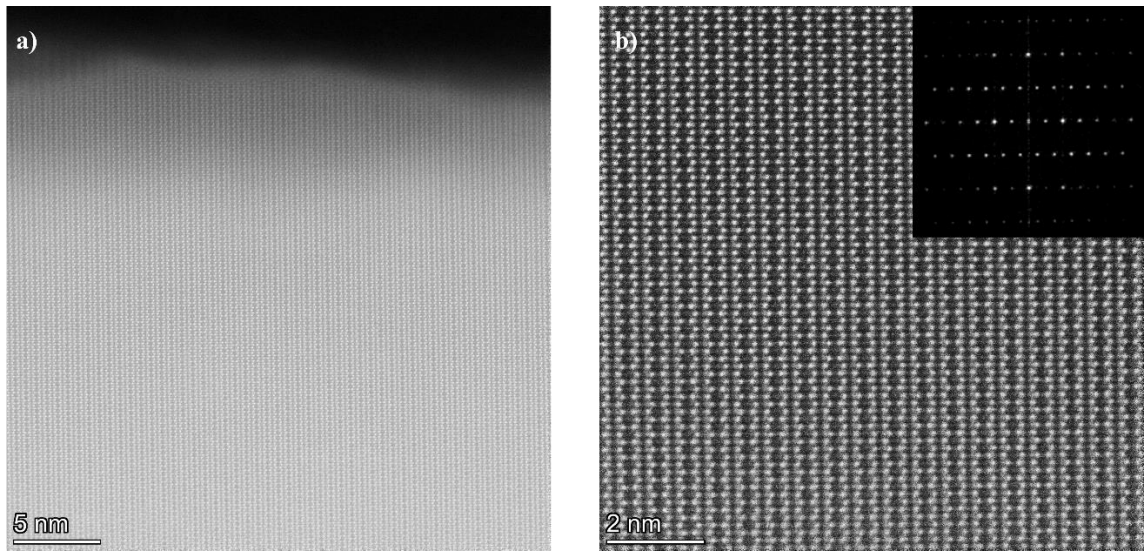

FIG S8. Atomic resolution HAADF-STEM images of implanted and annealed film with two resolutions, showing recovery of $\beta$ structure after annealing with subset image showing FFT of recovered lattice.